\documentclass[onecolumn,floatfix,showpacs,aps,nofootinbib, showkeys]{revtex4}

\usepackage[dvips]{epsfig}
\usepackage[english]{babel}
\usepackage{bbm}
\usepackage{verbatim}
\usepackage{array}
\usepackage{amsmath}
\usepackage{multirow}
\usepackage{amsfonts}
\usepackage{amssymb}
\usepackage{hyperref}
\usepackage{leading}
\usepackage[dvipsnames]{xcolor}
\hypersetup{colorlinks=true,linkbordercolor=Blue,linkcolor=Blue, citecolor=Blue}
\usepackage{subeqnarray}
\usepackage{setspace}
\usepackage{indentfirst} 

\usepackage{gensymb}

\begin{document}

\title{A hint towards mass dimension one Flag-dipole spinors}

\author{R. J. Bueno Rogerio$^{1}$} \email{rodolforogerio@unifei.edu.br}
\author{C. H. Coronado Villalobos$^{2}$} \email{carlos.coronado@inpe.br}
\author{A. R. Aguirre$^{1}$} \email{alexis.roaaguirre@unifei.edu.br}
\affiliation{$^{1}$Instituto de F\'isica e Qu\'imica, Universidade Federal de Itajub\'a - IFQ/UNIFEI, \\
Av. BPS 1303, CEP 37500-903, Itajub\'a - MG, Brazil.}
\affiliation{$^{2}$Instituto Nacional de Pesquisas Espaciais (INPE),\\
CEP 12227-010, S\~ao Jos\'e dos Campos - SP, Brazil.}


\begin{abstract}
{\textbf{Abstract.}}
In this report we advance in exploring further details concerning the formal aspects of the construction of a Flag-dipole spinor. We report a (re-)definition of the dual structure which provide a Lorentz invariant and non-null norm, ensuring a local theory. With the new dual structure at hands, we look towards define relevant physical amounts, e.g., spin sums and quantum field operator. As we will see, the Flag-dipole and the Elko's theory are quite familar. In this vein, it is possible, via a matrix \emph{transformation}, to write Flag-dipole spinors in terms of Elko spinor, evincing that both spinors are physically related and some physical amounts may be stated as equivalent.

\end{abstract}

\pacs{04.62.+v, 03.70.+k, 03.65.-w}
\keywords{Flag-dipole; Mass dimension one; dual helicity.}

\maketitle

\section{Introduction}\label{intro}

Spinors constitute a comprehensive mathematical object of study, with several applications in physics. In particular, such mathematical objects are the rudimentary ingredient to describe fermionic particles. Defined upon the Lorentz group, a lot have been looked at the space-time symmetries \cite{weinberg1,Wigner1,Wigner2}, underlying the Lounesto's classification \cite{lounestolivro}, rather than the symmetries of the spinor space itself \cite{julio2019} .

The Standard Model (SM) of particle physics take the Dirac, Majorana and Weyl spinors (and also the associated quantum field) as building blocks to describe the fermionic sector of the quantum field theory.
Quite recently, with the Elko --- mass dimension one fermions --- theoretical discovery, a new physics was brought to the light. Elko spinors were proposed in 2004 \cite{jcap}, the fermionic field of spin-${1/2}$ endowed with mass dimension one is constructed upon a complete set of eigenspinors of the charge conjugation operator. We point out that, contrary to what is commonly established in the seminal literature of Elko, about mass dimensionality one, different points of view of given subject are carried out on the works \cite{elkomonopoles,vazelko}, where the authors shows that in some very specific mathematical contexts, it is possible to show that Elko spinors satisfies a coupled system of first order partial differential equations for a pair of Elko and also a first order equation in the Clifford algebra of physical space, thus, in such context the authors claim the possibility of mass dimension $3/2$. 

The  features carried throughout mass dimension one theory,  opened windows to a physical content as known as \emph{beyond the Standard Model}, trying to answer many questions that seems to be incomplete (and perhaps unsolvable) until the present moment. We believe that a huge and interesting content is yet hidden beyond the mass dimension one fermions. 

In the present essay, we introduce new features concerning the Flag-dipole spinors recently proposed in \cite{dualtipo4}. As we will see, there is a strong connection between the Flag-dipole spinors and the Elko spinors, by defining some mathematical conditions it is possible to transmute among type-4 and Elko spinors, we emphasize that the last above mentioned spinors are shown to be slightly related to Elko just by a matrix transformation. All the results presented here, in such mathematical approach, were supported and also contribute with previous results found in \cite{chengflagdipole}. We highlight that the seminal attempt to stablish a connection among flag-dipole spinors and Elko, pointing out some algebraic and geometric considerations, was developed in \cite{da2010elko}.

This paper is organized as it follows: In section \ref{definicaotipo4} we make a brief overview on the Flag-dipole spinors definition. Section \ref{tipo4dual} is reserved to introduce a method to define the adjoint structure which furnish a Lorentz invariant and a non-null norm. Along this section, we also develop allusive comments showing the connection among Elko and Flag-dipole spinors. Thus, in section \ref{tipo4campo}, we explicit show the quantum field operator, where the expansion coefficients are Flag-dipole spinors written in terms of Elko spinors. Finally, in Sect. \ref{remarks} we conclude.       

\section{Elementary Review}\label{definicaotipo4}

This section is reserved for bookkeeping purposes. Here we shall describe the basic introductory elements concerning Flag-dipole spinors - highlighting the main and the necessary aspects that will be carried along the paper. 

\subsection{Setting up the notation}\label{conceitos}
To obtain an explicit form of a given $\psi(p^{\mu})$ spinor we first call out attention for the rest spinors, $\psi(k^{\mu})$. For an arbitrary momentum $(p^{\mu})$, we have the following condition
\begin{equation}\label{1}
\psi(p^{\mu}) = e^{i\kappa.\varphi}\psi(k^{\mu}),
\end{equation}
where the $\psi(k^{\mu})$ rest frame spinor is a direct sum of the $(1/2,0)$ and $(0,1/2)$ Weyl spinors, which are usually defined as 
\begin{equation}
\psi(k^{\mu}) = \left(\begin{array}{c}
\phi_R(k^{\mu}) \\ 
\phi_L(k^{\mu})
\end{array} \right).
\end{equation}
Note that  we define the $k^{\mu}$ rest frame momentum as 
\begin{equation}
k^{\mu}\stackrel{def}{=}\bigg(m,\; \lim_{p\rightarrow 0}\frac{\boldsymbol{p}}{p}\bigg), \; p=|\boldsymbol{p}|,
\end{equation}
and, by parametrizing the general four-momentum (in spherical coordinates) as 
\begin{equation}
p^{\mu}=(E, p\sin\theta\cos\phi, p\sin\theta\sin\phi, p\cos\theta),
\end{equation}
then, the boost operator is given as follows
\begin{eqnarray}\label{boostoperator}
e^{i\kappa.\varphi} = \sqrt{\frac{E+m}{2m}}\left(\begin{array}{cc}
\mathbbm{1}+\frac{\vec{\sigma}.\hat{p}}{E+m} & 0 \\ 
0 & \mathbbm{1}-\frac{\vec{\sigma}.\hat{p}}{E+m}
\end{array} \right),
\end{eqnarray}
where $\cosh\varphi = E/m, \sinh\varphi=p/m$, and  $\hat{\boldsymbol{\varphi}} = \hat{\boldsymbol{p}}$.
Thus, such momentum parametrization allow us to defined the right-hand and left-hand components in the rest frame. Under inspection of the helicity operator it directly provide
\begin{equation}
\vec{\sigma}\cdot\hat{p}\; \phi^{\pm}(k^{\mu}) = \pm \phi^{\pm}(k^{\mu}).
\end{equation}
Thus, the positive helicity component is given by 
\begin{equation}
\phi^{+}(k^{\mu}) = \sqrt{m}e^{ i\vartheta_{1}}\left(\begin{array}{c}
\cos(\theta/2)e^{-i\phi/2} \\ 
\sin(\theta/2)e^{i\phi/2}
\end{array}\right), 
\end{equation} 
and the negative helicity reads\footnote{We reserve the right to omit the right-hand or left-hand component label, as commonly presented in the textbooks, due to the fact that this title comes from the way such components are transformed by Lorentz transformations.}
\begin{equation}
\phi^{-}(k^{\mu}) = \sqrt{m}e^{ i\vartheta_{2}}\left(\begin{array}{c}
\sin(\theta/2)e^{-i\phi/2} \\ 
-\cos(\theta/2)e^{i\phi/2}
\end{array}\right).
\end{equation} 

As remarked in Ref \cite{mdobook}, the presence of the phase factor becomes necessary to set up the framework of eigenpinors of parity, or charge conjugation operators. One can verify that under a rotation by an angle $\vartheta$ the Dirac spinors pick up a global phase $e^{\pm i\vartheta/2}$, depending on the related helicity. However, this only happens for eigenspinors of parity operator. For the eigenspinors of charge conjugation operator, the phases factor must be $\vartheta_1=0$ and $\vartheta_2=\pi$ \cite{aaca}. Thus, this judicious phase combination ensure for the field the character of locality. Further details can be found in \cite{mdobook}.

\subsection{ A brief overview on Flag-dipole spinors}

The device developed in \cite{dualtipo4} to define the Flag-dipole spinors is, thus, based on defining a mapping matrix, which authors call by $M$. Such a matrix acts on the spinor's components and transmute a single helicity spinor into a dual helicity spinor and {vice-versa}. The mapping protocol is directly obtained if one employ the following conditions 
\begin{eqnarray}
M^{L}\psi_{\{+,+\}}(p^{\mu})= \Lambda_{\{+,-\}}(p^{\mu}), \label{lamb1}
\\
M^{R}\psi_{\{+,+\}}(p^{\mu})= \Lambda_{\{-,+\}}(p^{\mu}), \label{lamb2}
\\
M^{R}\psi_{\{-,-\}}(p^{\mu})= \Lambda_{\{+,-\}}(p^{\mu}), \label{lamb3}
\\
M^{L}\psi_{\{-,-\}}(p^{\mu})= \Lambda_{\{-,+\}}(p^{\mu}), \label{lamb4}
\end{eqnarray}
where the above single-helicity spinors $\psi(p^{\mu})$ are defined in \cite{diracpauli,nondirac}. To ensure the same result in all reference frames, the $M$ matrix must to commute with the boost operators given in \eqref{boostoperator}.

A careful inspection says that the relations \eqref{lamb1} and \eqref{lamb3} provides the same  spinor $\Lambda$, up to a global sign, and the same akin reasoning should be inferred to \eqref{lamb2} and \eqref{lamb4}. Said that, the mapping procedure starts by imposing that a positive-eigenvalue component transmute to a negative-eigenvalue component,  and vice-versa. Thus, such statement can be formally expressed as follows,
\begin{eqnarray}
\vec{\sigma}\cdot\hat{p}\;\mathcal{O}_{(+-)}\phi^{+}(k^{\mu}) = -\phi^{-}(k^{\mu}),
\end{eqnarray}
and
\begin{eqnarray}
\vec{\sigma}\cdot\hat{p}\;\mathcal{O}_{(-+)}\phi^{-}(k^{\mu}) = +\phi^{+}(k^{\mu}).
\end{eqnarray}
Note that the $\mathcal{O}$ operator is a $2\times 2$ matrix which is responsible to interchange the component's helicity. It is worth pointing out that the lower index ($+,-$) and ($-,+$) indicates that the operator $\mathcal{O}$ takes a component which carry positive eigenvalue to one which carry negative one, and vice-versa. The introduced operator composes a more general one, which will be denoted by $M$. Now, such more involved operator act over spinors, and transmute its helicities components. 
The $M$ matrix may be defined as follows 
\begin{equation}
M^{L} = \left(\begin{array}{cc}
\mathbbm{1} & 0 \\ 
0 & \mathcal{O}
\end{array} \right),
\end{equation}
which keeps the right-hand helicity unchanged and acts only on the right-hand component, and without loss of generality it may be defined
\begin{equation}
M^{R} = \left(\begin{array}{cc}
\mathcal{O} & 0 \\ 
0 & \mathbbm{1}
\end{array} \right),
\end{equation}
acting only on the right-hand component. Here $\mathbbm{1}$ stand for $2\times 2$ identity matrix.

A straight computation of such operators, allow one to explicitly defined 
\begin{equation}
M^{L}_{(+-)} = \left(\begin{array}{cccc}
1 & 0 & 0 & 0 \\ 
0 & 1 & 0 & 0 \\ 
0 & 0 & 0 & -e^{-i\phi} \\ 
0 & 0 & e^{i\phi} & 0
\end{array}\right), \quad M^{L}_{(-+)} = \left(\begin{array}{cccc}
1 & 0 & 0 & 0 \\ 
0 & 1 & 0 & 0 \\ 
0 & 0 & 0 & e^{-i\phi} \\ 
0 & 0 & -e^{i\phi} & 0
\end{array}\right), \label{ML17}
\end{equation}
and,
\begin{equation}
 M^{R}_{(+-)} = \left(\begin{array}{cccc}
0 & -e^{-i\phi} & 0 & 0 \\ 
e^{i\phi} & 0 & 0 & 0 \\ 
0 & 0 & 1 & 0 \\ 
0 & 0 & 0 & 1
\end{array}\right), \quad M^{R}_{(-+)} = \left(\begin{array}{cccc}
0 & e^{-i\phi} & 0 & 0 \\ 
-e^{i\phi} & 0 & 0 & 0 \\ 
0 & 0 & 1 & 0 \\ 
0 & 0 & 0 & 1
\end{array}\right),\label{MR18}
\end{equation}
where the upper indices $R$ and $L$ stands for the specific component (right-hand or left-hand) where $M$ acts. It is also possible to transmute from a dual-helicity spinor to a single-helicity one, due to the fact that the matrices in \eqref{ML17} and \eqref{MR18} are invertible.
In this manner, the single-helicity spinors previously introduced in \cite{nondirac}, allow us to built the following set of dual-helicity spinors in the rest frame
\begin{equation}
\Lambda_{\{+,-\}}(k^{\mu})=\sqrt{m}\left(\begin{array}{c}
\alpha_{+}\cos(\theta/2)e^{-i\phi/2}\\
\alpha_{+}\sin(\theta/2)e^{i\phi/2}\\
-\beta_{+}\sin(\theta/2)e^{-i\phi/2}\\
\beta_{+}\cos(\theta/2)e^{i\phi/2}\\
\end{array}
\right), \quad \Lambda_{\{-,+\}}(k^{\mu})=\sqrt{m}\left(\begin{array}{c}
-\alpha_{-}\sin(\theta/2)e^{-i\phi/2}\\
\alpha_{-}\cos(\theta/2)e^{i\phi/2}\\
\beta_{-}\cos(\theta/2)e^{-i\phi/2}\\
\beta_{-}\sin(\theta/2)e^{i\phi/2}
\end{array}
\right),
\end{equation}
where $\alpha_{\pm}$ and $\beta_{\pm}$ constant parameters $\in\mathbb{C}$, to be further determined. Remarkably enough, we stress that after the mapping procedure, the $(0,1/2)$ and $(1/2,0)$ representation spaces automatically show to be connected via Wigner's time-reversal operator, $\Theta$, an Elko's quite familiar feature. Such aforementioned feature is an intrinsic characteristic of dual-helicity spinors, as stated in \cite{interplay}.

By taking the spinor from the rest frame to an arbitrary momentum frame, we can write
\begin{equation}
\Lambda_{\{\pm,\mp\}}(p^{\mu}) =  e^{i\kappa.\varphi}\Lambda_{\{\pm,\mp\}}(k^{\mu}),
\end{equation}
and then, we have the following\footnote{In order to leave the notation clear to the reader, we choose to defined the Lorentz boost parameters as $\mathcal{B}_{\pm}\equiv\sqrt{\frac{E+m}{2m}}\big(1\pm \frac{p}{E+m}\big)$.} 
\begin{equation}\label{spinor12}
\Lambda_{\{+,-\}}(p^{\mu})=\mathcal{B}_{+}\left(\begin{array}{c}
-\alpha_{+}\Theta\phi_L^{-*}(k^{\mu}) \\ 
\beta_+\phi_L^{-}(k^{\mu})
\end{array} \right), \;  \Lambda_{\{-,+\}}(p^{\mu})=\mathcal{B}_{-}\left(\begin{array}{c}
\alpha_{-}\Theta\phi_L^{+*}(k^{\mu}) \\ 
\beta_{-}\phi_L^{+}(k^{\mu})
\end{array} \right).
\end{equation}
The above set of spinors fits into the fourth class within Lounesto classification \cite{lounestolivro} if one constrain the parameters $\vert\alpha\vert^{2}\neq\vert\beta\vert^{2}$, and both modulus different from the unit. We notice that, if the modulus of the phases are equal, or equal to the unit, then such spinors will fit into the fifth class within Lounesto's classification \cite{whereareelko, hoffdirac}. Such results are in agreement with what it was stated in \cite{rodolfoconstraints,interplay} - singular spinors must carry dual helicity feature and not necessarily hold conjugacy under charge conjugation operator. 

At this point we shall see the first strong affinity among Flag-dipole spinors and Elko spinors. It is possible to reconstruct Elko spinors from \eqref{spinor12}. Given task is easily accomplished by a convenient choice of phases, e.g., $\alpha=\pm i$ and $\beta=1$, holding the relation $\vert\alpha\vert^{2}=\vert\beta\vert^{2}$ \cite{dualtipo4}. To make the distinction between flag-pole and flag-dipole spinors clear to the reader, we reserved an Appendix with the explicit form of the bilinear amounts.  

\section{On the set-up of the adjoint structure}\label{tipo4dual}
As soon as we have explicitly defined the Flag-dipole spinors in the previous section, now we turn our attention to define the corresponding spinorial dual structure. So far, to be classified as a singular spinor within Lounesto's classification, the norm for Flag-dipole spinors stand null $(\sigma=0)$. Thus, in this section we will perform a mathematical program looking towards obtain a real and invariant norm under Lorentz transformations. 

Thus, we can compute the following
\begin{eqnarray}
\bar{\Lambda}_{\{+,-\}}(p^{\mu})\Lambda_{\{-,+\}}(p^{\mu}) = \alpha_{-}\beta^*_{+}m+\alpha^*_{+}\beta_{-}m,
\end{eqnarray}
and by imposing both representation space on equal terms, we set $\alpha_{-}\beta^*_{+}=1$ (or $\alpha_{-}\beta^*_{+}=-1$) and $\alpha^*_{+}\beta_{-}=1$ (or $\alpha^*_{+}\beta_{-}=-1$). Such relations provide\footnote{Remarkably enough the introduced spinors carry some formal aspects as Elko do. Note that the relations presented in \eqref{ortoS} and \eqref{ortoA} differ from Elko only by a imaginary phase factor. Nonetheless, the adjoint structure also must to flip the helicity to hold a Lorentz invariant norm. One may follow the same prescription presented in \cite{mdobook} to compute the dual structure, however, the result hold the same as presented here.}
\begin{eqnarray}
&&\bar{\Lambda}^S_{\{\pm,\mp\}}(p^{\mu})\Lambda^S_{\{\mp,\pm\}}(p^{\mu}) = 2m,\label{ortoS}
\\
&&\bar{\Lambda}^A_{\{\pm,\mp\}}(p^{\mu})\Lambda^A_{\{\mp,\pm\}}(p^{\mu}) = -2m,\label{ortoA}
\end{eqnarray}
and stands null for the remaining cases.
As we will soon check given phase imposition lead to interesting results. The upper index $S$ and $A$ is just a way to make distinction between the signs presented in the above orthonormal relations. We also notice that the orthonormal relations are phase independent, and the only constraint from now on is $|\beta_{\pm}|^2 \neq 1$. Otherwise, we do not guarantee a Flag-dipole spinor, in agreement with \cite{chengflagdipole}.
Such phases ``\emph{fixing}'' provide a complete set of spinors and turns possible to rewrite the spinors \eqref{spinor12} in the following fashion
\begin{equation}
\Lambda^S_{\{+,-\}}(p^{\mu})=\mathcal{B}_{+}\left(\begin{array}{c}
-\beta^{*-1}_{-}\Theta\phi_L^{-*}(k^{\mu}) \\ 
\beta_+\phi_L^{-}(k^{\mu})
\end{array} \right),
\; \Lambda^S_{\{-,+\}}(p^{\mu})=\mathcal{B}_{-}\left(\begin{array}{c}
\beta^{*-1}_{+}\Theta\phi_L^{+*}(k^{\mu}) \\ 
\beta_-\phi_L^{+}(k^{\mu})
\end{array} \right),
\end{equation} 
and
\begin{equation}
\Lambda^A_{\{+,-\}}(p^{\mu})=\mathcal{B}_{+}\left(\begin{array}{c}
\beta^{*-1}_{-}\Theta\phi_L^{-*}(k^{\mu}) \\ 
\beta_+\phi_L^{-}(k^{\mu})
\end{array} \right),
\; \Lambda^A_{\{-,+\}}(p^{\mu})=\mathcal{B}_{-}\left(\begin{array}{c}
-\beta^{*-1}_{+}\Theta\phi_L^{+*}(k^{\mu}) \\ 
\beta_-\phi_L^{+}(k^{\mu})
\end{array} \right).
\end{equation} 
With the previous results at hands, we are able to defined the dual as
\begin{eqnarray}\label{dual1}
\widetilde\Lambda^{S/A}_{\{+,-\}}(p^{\mu}) = \bar{\Lambda}^{S/A}_{\{-,+\}}(p^{\mu}), \quad
\widetilde\Lambda^{S/A}_{\{-,+\}}(p^{\mu}) = \bar{\Lambda}^{S/A}_{\{+,-\}}(p^{\mu}), 
\end{eqnarray} 
where bar means the Dirac adjoint structure ($\bar{\psi} = \psi^{\dag}\gamma_{0}$). Previous results lead to the following spin sums
\begin{equation}\label{somaS}
\sum_{h = \{\pm,\mp\}}\Lambda^{S}_{h}(p^{\mu})\widetilde\Lambda^{S}_{h}(p^{\mu}) = m[\mathbbm{1} + \mathcal{N}(\phi, \beta_{\pm}, \theta)],
\end{equation}
and 
\begin{equation}\label{somaA}
\sum_{h = \{\pm,\mp\}}\Lambda^{A}_{h}(p^{\mu})\widetilde\Lambda^{A}_{h}(p^{\mu}) = -m[\mathbbm{1} - \mathcal{N}(\phi,\beta_{\pm},\theta)],
\end{equation}
where the $\mathcal{N}(\phi, \beta_{\pm}, \theta)$ matrix reads 
\begin{eqnarray}\label{Nzao}
\mathcal{N}(\phi, \beta_{\pm}, \theta) = \left(\begin{array}{cccc}
0 & 0 & -\frac{\sin(\theta)}{|\beta_{-}|^2|\beta_{+}|^2}f(\beta_{+},\beta_{-}) & \frac{e^{-i\phi}}{|\beta_{-}|^2|\beta_{+}|^2}g(\beta_{+},\beta_{-}) \\ 
0 & 0 & \frac{e^{i\phi}}{|\beta_{-}|^2|\beta_{+}|^2}g^*(\beta_{+},\beta_{-}) & \frac{\sin(\theta)}{|\beta_{-}|^2|\beta_{+}|^2}f(\beta_{+},\beta_{-}) \\ 
-\sin(\theta)f(\beta_{+},\beta_{-}) & e^{-i\phi} g(\beta_{+},\beta_{-}) & 0 & 0 \\ 
e^{i\phi} g^*(\beta_{+},\beta_{-})& \sin(\theta)f(\beta_{+},\beta_{-}) & 0 & 0
\end{array} \right),
\end{eqnarray}
with $f(\beta_{+},\beta_{-})\equiv \beta^*_{-}\beta_{+}+\beta^*_{+}\beta_{-}$ and $g(\beta_{+},\beta_{-})\equiv [\cos^2(\theta/2)f(\beta_{+},\beta_{-}) - \beta^*_{-}\beta_{+}]$. We elucidate some interesting facts concerning the Flag-dipole spin sums: first, we have $\det[\mathbbm{1}\pm \mathcal{N}(\phi, \beta_{\pm},\theta)]=0$, i.e., such matrix do not hold inverse; second, note that $\mathcal{N}^2(\phi, \beta_{\pm}, \theta)=\mathbbm{1}$ and $\mathcal{N}^{-1}(\phi, \beta_{\pm}, \theta)=\mathcal{N}(\phi, \beta_{\pm}, \theta)$. Finally, a remarkably feature is that, under certain phase imposition, it is possible to recover the Elko's spin sums, i.e., $\mathcal{N}(\phi, \beta_{\pm}, \theta)\rightarrow \mathcal{G}(\phi)$. Now, notice
\begin{equation}
\mathcal{N}(\phi, \beta_{\pm}, \theta)\Lambda^{S}_{h}(p^{\mu}) = \Lambda^{S}_{h}(p^{\mu}),
\end{equation}
and
\begin{equation}
\mathcal{N}(\phi, \beta_{\pm}, \theta)\Lambda^{A}_{h}(p^{\mu}) = -\Lambda^{A}_{h}(p^{\mu}).
\end{equation}    
albeit the $\Lambda_{h}(p^{\mu})$ are annihilated by the operator $\mathcal{N}(\phi, \beta_{\pm}, \theta)$, it can not represent a wave-function due to the fact that it does not have time dependence.
It is worth mentioning that the Flag-dipole spin sums furnish the completeness relation 
\begin{equation}
\frac{1}{2m}\sum_{h= \{\pm,\mp\}}\big[\Lambda^{S}_{h}(p^{\mu})\widetilde\Lambda^{S}_{h}(p^{\mu}) - \Lambda^{A}_{h}(p^{\mu})\widetilde\Lambda^{A}_{h}(p^{\mu})\big] = \mathbbm{1},
\end{equation}
as expected. 

We emphasize the difficulty to physically interpret the amounts above, since without phases fixing, we are in a totally arbitrary framework. Note that there is a impossibility to fix the phases via discrete symmetries or even by dynamic.
In view of this facts, locality seems to be affected with the previously term presented in the spin sums. We are looking towards a method to bypass such a problem and bring some light to the interpretation.  

\subsection{A guide for a new adjoint}

So far we have defined the dual structure for the Flag-dipole spinors, introduced in \eqref{dual1}, an element which may bring serious difficulties to the physical interpretation has emerged, check for the right-hand side of Eq.\eqref{Nzao}. As one can see, $\mathcal{N}(\phi, \beta, \theta)$ is not Lorentz invariant due to $\phi$ and $\theta$ dependence. If we proceed with the quantum field operator and the quantum correlators calculations, such term will appear, forcing us to face a non-local theory. 

The mathematical device to be used here is similar to the one previously developed in Refs. \cite{aaca,vicinity}, both supported by \cite{barata}. Thus, we define the following dual structures
\begin{eqnarray}
\widetilde{\Lambda}^{S}_h(p^\mu) &\rightarrow& \widetilde{\Lambda}\;^{S}_h(p^\mu)\mathcal{D}_{S}, \\
\widetilde{\Lambda}^{A}_h(p^\mu) &\rightarrow &\widetilde{\Lambda}\;^{A}_h(p^\mu)\mathcal{D}_{A},
\end{eqnarray}  
where the operators $\mathcal{D}_{S}$ and $\mathcal{D}_{A}$ require some important properties, namely, the spinors $\Lambda^{S}_h(p^\mu)$ and $\Lambda^{A}_h(p^\mu)$ must be eigenspinors of $\mathcal{D}_{S}$ and $\mathcal{D}_{A}$ respectively, with eigenvalues given by the unity, ensuring the following relations
\begin{eqnarray}\label{7}
\mathcal{D}_{S}\Lambda^{S}_h(p^\mu) = \Lambda^{S}_h(p^\mu), \quad\quad \mathcal{D}_{A}\Lambda^{A}_h(p^\mu)=\Lambda^{A}_h(p^\mu).
\end{eqnarray} 
Besides, both operators also must fulfill 
\begin{eqnarray}\label{8}
\widetilde{\Lambda}^{S}_h(p^\mu)\mathcal{D}_{S}\Lambda^{A}_h(p^\mu) = 0, \quad\quad \widetilde{\Lambda}^{A}_h(p^\mu)\mathcal{D}_{A}\Lambda^{S}_h(p^\mu) = 0,
\end{eqnarray}  
in order to satisfy and keep unchanged the orthonormal relations \eqref{ortoS} and \eqref{ortoA}.
Following the same prescription as in \cite{aaca}, one found the characteristic polynomial to be
\begin{equation}\label{polinomio}
p(\upsilon) = \upsilon^4 - 2\upsilon^2 + 1,
\end{equation}
and then, the $\mathcal{N}(\phi, \beta, \theta)$ hold the following set of eigenvalues $\pm 1$ each one with multiplicity two. In view of the above considerations, we reach to the definition of the $\mathcal{D}$ operators, which read
\begin{eqnarray}
\mathcal{D}_{S} =  2\bigg[\frac{\mathbbm{1}-\tau\mathcal{N}(\phi, \beta, \theta)}{1-\tau^2}\bigg], 
\\
\mathcal{D}_{A} = 2\bigg[\frac{\mathbbm{1}+\tau\mathcal{N}(\phi, \beta, \theta)}{1-\tau^2}\bigg],
\end{eqnarray}
in which the parameter $\tau$ is responsible to provide the right inverse of the spin sum, exactly on the limit $\tau\rightarrow 1$, given by the roots of the relation \eqref{polinomio}.

Hence, the spin sums now read
\begin{eqnarray}
\label{spinsumsA1}\sum_{\alpha} \Lambda^S_{\alpha}(p^{\mu})\widetilde{\Lambda}^{S}_{\alpha}(p^{\mu})&=& 2m[\mathbbm{1}+\mathcal{N}(\phi, \beta, \theta)]\bigg[\frac{\mathbbm{1}-\tau\mathcal{N}(\phi, \beta, \theta)}{1-\tau^2}\bigg]\Big\vert_{\tau\rightarrow 1}\nonumber\\ 
&=&2m\mathbbm{1},
\\
\label{spinsumaA2}\sum_{\alpha} \Lambda^A_{\alpha}(p^{\mu})\widetilde{\Lambda}^{A}_{\alpha}(p^{\mu})&=& -2m[\mathbbm{1}-\mathcal{N}(\phi, \beta, \theta)]\bigg[\frac{\mathbbm{1}+\tau\mathcal{N}(\phi, \beta, \theta)}{1-\tau^2}\bigg]\Big\vert_{\tau\rightarrow 1}\nonumber\\
&=&-2m\mathbbm{1}.
\end{eqnarray}
Note that the general protocol developed here is analogous to the Elko's adjoint redefinition. We suppose that such similarities comes from the fact that both theories carry objects endowed with mass dimension one and dual helicity.

\subsection{Some constraints on the phases and a strong connection between Flag-dipole and Elko spinors}
In view of the results presented in \cite{chengflagdipole}, if one impose that $\beta_{-}=-iz$ and $\beta_{+}=z$, with $|z|^2\neq 1$, then we reach to the following results
 \begin{eqnarray}
&&\Lambda^{S}_{\{+,-\}}(p^{\mu}) = \mathcal{Z}(z)\lambda^S_{\{+,-\}}(p^{\mu}), \\
&&\Lambda^{S}_{\{-,+\}}(p^{\mu}) = -i\mathcal{Z}(z)\lambda^S_{\{-,+\}}(p^{\mu}), \\
&&\Lambda^{A}_{\{+,-\}}(p^{\mu}) = \mathcal{Z}(z)\lambda^A_{\{+,-\}}(p^{\mu}), \\
&&\Lambda^{A}_{\{-,+\}}(p^{\mu}) = -i\mathcal{Z}(z)\lambda^A_{\{-,+\}}(p^{\mu}),
 \end{eqnarray}
with $\mathcal{Z}(z)$ being a dimensionless matrix described by
\begin{equation}
\mathcal{Z}(z) = \left(\begin{array}{cc}
z^{*-1}\mathbbm{1} & 0 \\ 
0 & z\mathbbm{1}
\end{array}  \right).
\end{equation}  
Now, the associated adjoint structure written in terms of the Elko spinors, reads 
\begin{eqnarray}
&&\widetilde{\Lambda}^{S}_{\{+,-\}}(p^{\mu}) = \stackrel{\neg}{\lambda}^{S}_{\{+,-\}}(p^{\mu})\mathcal{Z}^{-1}(z)\mathcal{A}_{z}, \\
&&\widetilde{\Lambda}^{S}_{\{-,+\}}(p^{\mu}) = i\stackrel{\neg}{\lambda}^{S}_{\{-,+\}}(p^{\mu})\mathcal{Z}^{-1}(z)\mathcal{A}_{z}, \\
&&\widetilde{\Lambda}^{A}_{\{+,-\}}(p^{\mu}) = \stackrel{\neg}{\lambda}^{A}_{\{+,-\}}(p^{\mu})\mathcal{Z}^{-1}(z)\mathcal{B}_{z}, \\
&&\widetilde{\Lambda}^{A}_{\{-,+\}}(p^{\mu}) = i\stackrel{\neg}{\lambda}^{A}_{\{-,+\}}(p^{\mu})\mathcal{Z}^{-1}(z)\mathcal{B}_{z}.
\end{eqnarray}
The above relations hold the same result introduced in \eqref{ortoS} and \eqref{ortoA}, in the $\tau-$limit, but now replacing the Flag-dipole mass $m$ by the Elko's mass $m_{\lambda}$. Also, note that the operators $\mathcal{A}$ and $\mathcal{B}$ are $z$ dependent.

Although Flag-dipole spinors can be written in terms of Elko, it does not necessarily hold conjugacy under $\mathcal{C}=\gamma_2 \mathcal{K}$, as highlighted in \cite{interplay}, 
\begin{eqnarray}
&&\mathcal{C}\Lambda^{S}_{\{+,-\}}(p^{\mu}) = \mathcal{Z}^{-1}(z)\lambda^S_{\{+,-\}}(p^{\mu}), \\
&&\mathcal{C}\Lambda^{S}_{\{-,+\}}(p^{\mu}) = i\mathcal{Z}^{-1}(z)\lambda^S_{\{-,+\}}(p^{\mu}), \\
&&\mathcal{C}\Lambda^{A}_{\{+,-\}}(p^{\mu}) = \mathcal{Z}^{-1}(z)\lambda^A_{\{+,-\}}(p^{\mu}), \\
&&\mathcal{C}\Lambda^{A}_{\{-,+\}}(p^{\mu}) = i\mathcal{Z}^{-1}(z)\lambda^A_{\{-,+\}}(p^{\mu}),
 \end{eqnarray}
in agreement with what it was stated about Flag-dipole spinors and charge conjugation in Ref \cite{interplay}. Nonetheless, if one act twice with charge conjugation operator is expected to obtain $\mathcal{C}^2=\mathbbm{1}$. The non-neutral behaviour under $C$, allow one to note that interactions with the electromagnetic field should not be discarded. It is worth mentioning when cosmological applications are performed with this new field, we expect great similarities to those of the usual Elko field. However, due to its possibility of direct coupling with the electromagnetic field and other particles of the Standard Model, the reheating phase of the universe need to transfer energy to the baryonic components of the universe is easier to perform than with the Elko field.

A parenthetic remark related with the behaviour under action of charge conjugation operator is highlighted here. The expected interactions of the such fermions are not only restricted to dimension-four quartic self-interaction, and a dimension-four coupling with Higgs boson. Different from Elko spinors, the Flag-dipole opens up the possibility that the new field provides interaction with more particles of the Standard Model. 

\section{An Attempt to define the Flag-dipole quantum field}\label{tipo4campo}
In this section we use the $\Lambda(p^{\mu})$ spinors, introduced in the previous sections, to play the role of expansion coefficients of a Flag-dipole quantum field. Thus, it reads
\begin{eqnarray}
\mathfrak{f}_{\Lambda}(x) = \int\frac{d^3p}{(2\pi)^3}\frac{1}{\sqrt{2m_{\Lambda}E_{\Lambda}(\boldsymbol{p})}}\sum_{h}\big[a_h(\boldsymbol{p})\Lambda^S_{h}(\boldsymbol{p})e^{-ip_{\mu}x^{\mu}}
+b_h^{\dag}(\boldsymbol{p})\Lambda^A_{h}(\boldsymbol{p})e^{ip_{\mu}x^{\mu}}
\big],
\end{eqnarray}
 and its associated dual quantum field
 \begin{eqnarray}
\widetilde{\mathfrak{f}}_{\Lambda}(x) = \int\frac{d^3p}{(2\pi)^3}\frac{1}{\sqrt{2m_{\Lambda}E_{\Lambda}(\boldsymbol{p})}}\sum_{h}\big[a_h^{\dag}(\boldsymbol{p})\widetilde{\Lambda}^S_{h}(\boldsymbol{p})e^{ip_{\mu}x^{\mu}}
+b_h(\boldsymbol{p})\widetilde{\Lambda}^A_{h}(\boldsymbol{p})e^{-ip_{\mu}x^{\mu}}
\big],
\end{eqnarray}
in which $\Lambda^S(\boldsymbol{p})$ and $\Lambda^A(\boldsymbol{p})$ are the positive and negative energy spinors, respectively, and $m_{\Lambda}$ and $E_{\Lambda}(\boldsymbol{p})$ stands for the Flag-dipole mass and energy. 
The operators $a_h(\boldsymbol{p})$ and $b_h(\boldsymbol{p})$ are noted here as distinct objects. Such operators obey the anti-commutative fermionic relations
\begin{eqnarray}
&&\lbrace a_{h}(\boldsymbol{p}), a^{\dagger}_{h^\prime}(\boldsymbol{p^\prime})\rbrace = \lbrace b_{h}(\boldsymbol{p}), b^{\dagger}_{h^\prime}(\boldsymbol{p^\prime})\rbrace =(2\pi)^3\delta^3(\boldsymbol{p} - \boldsymbol{p^\prime})\delta_{hh^\prime},
\\
&&\lbrace a^{\dagger}_{h}(\boldsymbol{p}), a^{\dagger}_{h^\prime}(\boldsymbol{p^\prime})\rbrace = \lbrace a_{h}(\boldsymbol{p}), a_{h^\prime}(\boldsymbol{p^\prime})\rbrace = 0,
\\
&&\lbrace b^{\dagger}_{h}(\boldsymbol{p}), b^{\dagger}_{h^\prime}(\boldsymbol{p^\prime})\rbrace = \lbrace b_{h}(\boldsymbol{p}), b_{h^\prime}(\boldsymbol{p^\prime})\rbrace = 0.
\end{eqnarray} 

In order to ensure Fermi statistics, we make use of the above relations and also of the redefined adjoint structure. After some algebra, one get the following relations\footnote{We emphasize that if we instead work with the first dual formulation, then the equivalent anti-commutator would be
 \begin{equation}
\Big\lbrace \mathfrak{f}_{\Lambda}(\boldsymbol{x},t), \frac{\partial\widetilde{\mathfrak{f}}_{\Lambda}(\boldsymbol{x}^{\prime},t)}{\partial t}\Big\rbrace = i\delta^{3}(\boldsymbol{x} - \boldsymbol{x}^{\prime})+i\int \frac{d^3 p}{(2\pi)^3}e^{i\boldsymbol{p}(\boldsymbol{x}-\boldsymbol{x^{\prime}})}\mathcal{N}(\phi,\beta_{\pm},\theta).
\end{equation}
Where, up to our knowledge, the integration in $\mathcal{N}(\phi,\beta,\theta)$ is quite difficult and impracticable, bringing several difficulties to the physical interpretation.}
 \begin{eqnarray}
&& \Big\lbrace \mathfrak{f}_{\Lambda}(\boldsymbol{x},t), \frac{\partial\widetilde{\mathfrak{f}}_{\Lambda}(\boldsymbol{x}^{\prime},t)}{\partial t}\Big\rbrace\Big\vert_{\tau\rightarrow 1} = i\delta^{3}(\boldsymbol{x} - \boldsymbol{x}^{\prime}), \label{comutadorlocal}
\\
&&\lbrace \mathfrak{f}_{\Lambda}(\boldsymbol{x},t), \widetilde{\mathfrak{f}}_{\Lambda}(\boldsymbol{x}^{\prime},t)\rbrace=  0, 
\end{eqnarray}

Interesting enough, if we write the above amounts in terms of the Elko spinors, as a main outcome we have
\begin{eqnarray}
\mathfrak{f}_{\Lambda}(x) = \mathcal{Z}(z)\int\frac{d^3p}{(2\pi)^3}\frac{1}{\sqrt{2m_{\Lambda}E_{\Lambda}(\boldsymbol{p})}}\big[\big(a_{\{+,-\}}(\boldsymbol{p})\lambda^S_{\{+,-\}}(\boldsymbol{p})-ia_{\{-,+\}}(\boldsymbol{p})\lambda^S_{\{-,+\}}(\boldsymbol{p})\big)e^{-ip_{\mu}x^{\mu}}+
\nonumber\\
\big(b_{\{+,-\}}^{\dag}(\boldsymbol{p})\lambda^A_{\{+,-\}}(\boldsymbol{p})-ib_{\{-,+\}}^{\dag}(\boldsymbol{p})\lambda^A_{\{-,+\}}(\boldsymbol{p})\big)e^{ip_{\mu}x^{\mu}}
\big],
\end{eqnarray} 
and the adjoint field is given by
\begin{eqnarray}
\widetilde{\mathfrak{f}}_{\Lambda}(x) = \mathcal{Z}^{\dag}(z)\int\frac{d^3p}{(2\pi)^3}\frac{1}{\sqrt{2m_{\Lambda}E_{\Lambda}(\boldsymbol{p})}}\big[\big(a^{\dag}_{\{+,-\}}(\boldsymbol{p})\stackrel{\neg}{\lambda}^S_{\{+,-\}}(\boldsymbol{p})+ia^{\dag}_{\{-,+\}}(\boldsymbol{p})\stackrel{\neg}{\lambda}^S_{\{-,+\}}(\boldsymbol{p})\big)e^{ip_{\mu}x^{\mu}}+
\nonumber\\
\big(b_{\{+,-\}}^{\dag}(\boldsymbol{p})\stackrel{\neg}{\lambda}^A_{\{+,-\}}(\boldsymbol{p})+ib_{\{-,+\}}^{\dag}(\boldsymbol{p})\stackrel{\neg}{\lambda}^A_{\{-,+\}}(\boldsymbol{p})\big)e^{-ip_{\mu}x^{\mu}}
\big].
\end{eqnarray}
As soon as we define the quantum field for the Flag-dipole spinors, we have the following relations (in the $\tau-$limit)
\begin{equation}\label{correlatorelkoflag}
 \Big\lbrace \mathfrak{f}_{\Lambda}(\boldsymbol{x},t), \frac{\partial\widetilde{\mathfrak{f}}_{\Lambda}(\boldsymbol{x}^{\prime},t)}{\partial t}\Big\rbrace \propto \mathcal{Z}(z)\mathcal{Z}^{\dag}(z)\frac{m_{\lambda}}{m_{\Lambda}}\; i\delta^{3}(\boldsymbol{x} - \boldsymbol{x}^{\prime}), 
\end{equation}
in which
\begin{eqnarray}
\mathcal{Z}(z)\mathcal{Z}^{\dag}(z) = \left(\begin{array}{cc}
|z|^{-2}\mathbbm{1} & 0 \\ 
0 & |z|^{2}\mathbbm{1}
\end{array} \right).
\end{eqnarray}
Once again we emphasize that if $|z|^{2} = |z|^{-2} =1$, and consequently $\mathcal{Z}(z)\mathcal{Z}^{\dag}(z)=\mathbbm{1}$, then we recover the usual Elko's case. Although the extra dimensionless term appearing on the right-hand side of \eqref{correlatorelkoflag}, it does not bring any constraint about locality. The other quantum correlator reads
\begin{equation}
\lbrace \mathfrak{f}_{\Lambda}(\boldsymbol{x},t), \widetilde{\mathfrak{f}}_{\Lambda}(\boldsymbol{x}^{\prime},t)\rbrace=  0. 
\end{equation}
Note that such a results are only possible after define the Flag-dipole spinors in terms of the Elko spinors. Otherwise, elements of non-locality may arise in the correlators above. 

As a consequence shown in \cite{chengflagdipole}, we find the mass dimensionality of the quantum fields above to be one, and then its free lagrangian can be written as 
\begin{equation}\label{equa69}
\mathcal{L}_0 = \partial_{\mu}\widetilde{\mathfrak{f}}_{\Lambda}(x)\partial^{\mu}\mathfrak{f}_{\Lambda}(x) - m^2\widetilde{\mathfrak{f}}_{\Lambda}(x)\mathfrak{f}_{\Lambda}(x).
\end{equation}
 At this point, we should make some comments concerning possible interactions terms. As we shall see, the Flag-dipole spinors do not hold conjugacy under $\mathcal{C}$. In view of this fact, interactions with the electromagnetic field are allowed. Thus, Flag-dipole may carry self-interaction and also interaction with others particles of the Standard Model.
The only restriction is driven by the argument of power counting. 

In particular, if we consider local transformation on the $\Lambda(p^{\mu})$ spinors of the following way,
\begin{equation}
     \Lambda(p^{\mu}) \to \Lambda'(p^{\mu}) = e^{i\alpha G} \Lambda(p^{\mu}), 
\end{equation}
where $\alpha$ is a real parameter, and $G^{\dagger}=G$ is a $4\times 4$ Hermitian matrix, then in order to leave the bilinear terms in the lagrangian  $\mathcal{L}_0$ invariant under this global transformation, the matrix $G$ has to satisfy that $\left[G,\gamma_0\right]=0.$ We note that the matrices $\mathbbm{1}, \gamma_0$, and $\gamma_5\gamma_k$ (with $k=1, 2, 3$) are trivial solutions of that constraint. The case of the identity matrix lead us to the possibility of having exactly the $U(1)$ interactions.

\section{Concluding Remarks}\label{remarks}
In this communication we have developed a deeper analysis of the recently proposed Flag-dipole spinors in Ref \cite{dualtipo4}, by providing new features, as the definition of a dual structure, which leads to a Lorentz invariant and non-null norm. As it was shown, we have faced a difficulty in fixing the phase parameters due to the fact that Flag-dipole spinors do not compose a complete set of eigenspinors of discrete symmetries. That aforementioned fact forbid a complete phase fixing. However, it was possible to reduce from four to two arbitrary phases, providing the constraint $|\beta_{\pm}|^2\neq 1$. 

Nonetheless, we are able to compute the spin sums in a Flag-dipole general framework even with an arbitrary phase parameter open. Clearly, as it can be seen, the spin sums carry an element which is not manifestly invariant (or covariant) under Lorentz transformations. Interesting enough, we call attention to the fact that such an element that appears on the spin sums hold the same features as the one that appeared in the first Elko's formulation, namely the $\mathcal{G}(\phi)$ matrix. Even in the arbitrary phases set-up, we have shown that the completeness relation provide an expected result, being proportional to the identity. Going further, we have provided a mathematical mechanism to redefine the adjoint structure, in order to overcome the non-local term presented in the spin sums.      

A remarkable fact lies in the existing similarity between the methods presented in \cite{chengflagdipole} and \cite{dualtipo4} to obtain the Flag-dipole spinors. We shall finalize making an allusive assertion, by highlighting the similarities among Flag-dipole spinors and Elko spinors, supported by \cite{chengflagdipole}. We have performed a similar prescription to the case at hands, and as a result we have also shown the possibility of establishing a relation between Flag-dipole spinors and Elko spinors and \emph{vice-versa}. However, even if it is possible to write Flag-dipole spinors in terms of Elko, it does not hold conjugacy under charge-conjugation operator, which open windows to a variety of interactions with particles of the Standard Model. The procedure allow us to write the Flag-dipole adjoint structure, spin sums and the quantum field operators in terms of the previously Elko's last mentioned amounts. Such strong connection with the Elko provides a hint towards mass dimensionality and locality of the Flag-dipole spinors.

\section{Acknowledgements}
Authors express their gratitude to Professor Saulo Henrique Pereira, for his insightful comments concerning to cosmological applications. We also express gratitude to the anonymous Referee, his/her comments significantly improved the manuscript. RJBR thanks CNPq Grant N$^{\circ}$. 155675/2018-4 for the financial support and CHCV thanks CNPq PCI Grant N$^{\circ}$. 300236/2019-0 for the financial support. ARA thanks CAPES for partial financial support.

\appendix 
\section{Singular spinor bilinear forms}
This section is reserved to explicitly show the expressions for the bilinear covariants in terms of the parameters of the given singular spinor. Let $\lambda$ to be a singular spinor, thus, it reads
\begin{eqnarray}
\lambda =  \sqrt{m}\left( \begin{array}{c} \alpha\cos \left( {\it \theta}/2 \right) {{\rm e}^
{-i/2{\it \phi}}}\\ \noalign{\medskip}\alpha\sin \left( {\it \theta}/2
 \right) {{\rm e}^{i/2{\it \phi}}}\\ \noalign{\medskip}-\beta\sin \left( {
\it \theta}/2 \right) {{\rm e}^{-i/2{\it \phi}}}\\ \noalign{\medskip}\beta
\cos \left( {\it \theta}/2 \right) {{\rm e}^{i/2{\it \phi}}}\end{array}
 \right),
\end{eqnarray}
and the corresponding dual structure ($\bar{\lambda}=\lambda^{\dag}\gamma_0$)
\begin{eqnarray}
\bar{\lambda} = \sqrt{m}\left( \begin{array}{cccc}
-\beta^{*}\sin(\theta/2)e^{i\phi/2} & \beta^{*}\cos(\theta/2)e^{-i\phi/2} & \alpha^{*}\cos(\theta/2)e^{i\phi/2} & \alpha^{*}\sin(\theta/2)e^{i\phi/2}
\end{array} \right).
\end{eqnarray}
Thus, the bilinear forms read
\begin{eqnarray}
&&\sigma = 0, 
\\
&&\omega = 0,
\end{eqnarray}
\begin{eqnarray}
&&J_0 = |\alpha|^2 m+|\beta|^2 m,
\\
&&J_1 = -\sin(\theta)\cos(\phi)[|\alpha|^2+|\beta|^2]m,
\\
&&J_2 = -\sin(\theta)\sin(\phi)[|\alpha|^2+|\beta|^2]m,
\\
&&J_3 = -2\cos(\theta/2)[|\alpha|^2+|\beta|^2]m+|\alpha|^2 m+|\beta|^2 m,
\end{eqnarray}
\begin{eqnarray}
&&K_0 = |\alpha|^2 m -|\beta|^2 m, \label{k0}
\\
&&K_1 = -\sin(\theta)\cos(\phi)[|\alpha|^2-|\beta|^2]m,
\\
&&K_2 = -\sin(\theta)\sin(\phi)[|\alpha|^2-|\beta|^2]m,
\\
&&K_3 = -2\cos(\theta/2)[|\alpha|^2-|\beta|^2]m+|\alpha|^2 m-|\beta|^2 m,\label{k3}
\end{eqnarray} 
and
\begin{eqnarray}
&&S_{01} = \frac{i}{2} [2\cos^2(\theta/2)\cos(\phi)(\alpha^{*}\beta-\alpha\beta^{*})-\alpha^{*}\beta e^{-i\phi}+\alpha\beta^{*} e^{i\phi}]m,
\\
&&S_{02} = \frac{1}{2}[2i\cos^2(\theta/2)\sin(\phi)(\alpha^{*}\beta-\alpha\beta^{*})+\alpha^{*}\beta e^{-i\phi}+\alpha\beta^{*} e^{i\phi}]m,
\\
&&S_{03} = -\frac{i}{2} \; \sin \left( {\it \theta}\right) \left( \alpha^{*}\beta-\beta^{*}\alpha\right)m,
\\
&&S_{12} = -\frac{1}{2}\;\sin(\theta) \left( \alpha^{*}\beta+\beta^{*}\alpha\right)m, 
\\
&&S_{13} = \frac{i}{2}[2i\cos^2(\theta/2)\sin(\phi)(\alpha^{*}\beta-\alpha\beta^{*})+\alpha^{*}\beta e^{-i\phi}-\alpha\beta^{*} e^{i\phi}]m,
\\
&&S_{23}= \frac{1}{2} [2\cos^2(\theta/2)\cos(\phi)(\alpha^{*}\beta+\alpha\beta^{*})-\alpha^{*}\beta e^{-i\phi}-\alpha\beta^{*} e^{i\phi}]m.
\end{eqnarray}
Note that, intrinsically, the distinction among flag-pole and flag-dipole spinors is associated with the moduli of the phases factors. 
\bibliographystyle{unsrt}
\bibliography{refs}

\end{document}